\documentclass[aps,pra,twocolumn,superscriptaddress,notitlepage,showpacs,showkeys]{revtex4-1}
\usepackage{graphicx,amsmath,amsfonts,amssymb,upgreek,txfonts,color}
\usepackage[colorlinks,linkcolor=blue,citecolor=blue,urlcolor=blue,breaklinks=true]{hyperref}
\usepackage[utf8x]{inputenc}

\begin{document}

\title{ Measurement of entropy and quantum coherence properties of two type-I entangled photonic qubits
	}

\author{Ali Motazedifard} 
\email{motazedifard.ali@gmail.com}
\address{Quantum Optics Group, Iranian Center for Quantum Technologies (ICQTs), Tehran, Iran}
\address{Quantum Communication Group, Iranian Center for Quantum Technologies (ICQTs), Tehran, Iran}
\address{Quantum Sensing and Metrology Group, Iranian Center for Quantum Technologies (ICQTs), Tehran, Iran}

\author{Seyed Ahmad Madani} 
\address{Quantum Optics Group, Iranian Center for Quantum Technologies (ICQTs), Tehran, Iran}
\address{Quantum Communication Group, Iranian Center for Quantum Technologies (ICQTs), Tehran, Iran}

\author{N. S. Vayaghan} 
\address{Quantum Optics Group, Iranian Center for Quantum Technologies (ICQTs), Tehran, Iran}
\address{Quantum Communication Group, Iranian Center for Quantum Technologies (ICQTs), Tehran, Iran}

\date{\today}
\begin{abstract}
Using the type-I SPDC process in BBO nonlinear crystal (NLC), we generate a polarization-entangled state near to the maximally-entangled Bell-state with high-visibility (high-brightness) $ 98.50 \pm 1.33 ~ \% $ ($ 87.71 \pm 4.45 ~  \% $) for HV (DA) basis. We calculate the CHSH version of the Bell inequality, as a nonlocal realism test, and find a strong violation from the classical physics or any hidden variable theory (HVT), $ S= 2.71 \pm 0.10 $. 
Via measuring the coincidence count (CC) rate in the SPDC process, we obtain the quantum efficiency of single-photon detectors (SPDs) around $ (25.5\pm 3.4) \%  $, which is in good agreement to their manufacturer company. As expected, we verify the linear dependency of the CC rate vs. pump power of input CW-laser, which may yield to find the effective second-order susceptibility crystal. 
Using the theory of the measurement of qubits, includes a tomographic reconstruction of quantum states due to the linear set of 16 polarization-measurement, together with a maximum-likelihood-technique (MLT), which is based on the numerical optimization, we calculate the physical non-negative definite density matrices, which implies on the non-separability and entanglement of prepared state. By having the maximum likelihood density operator, we calculate precisely the entanglement measures such as Concurrence, entanglement of formation, tangle, logarithmic negativity, and different entanglement entropies such as linear entropy, Von-Neumann entropy, and Renyi 2-entropy. Finally, this high-brightness and low-rate entangled photons source can be used for short-range quantum measurements in the Lab.

\end{abstract}


\maketitle

\section{Introduction \label{sec1}}

Creation, preparation, manipulation, and characterization of entangled quantum states, i.e., entanglement measures and density operator, are important in different areas of quantum technologies such as quantum computing, quantum cryptography, quantum communications, quantum metrology as well as the fundamentals of quantum information sciences. 
To date, the cleanest, easiest, and most inexpensive and accessible entanglement source, notably the polarization entanglement source, can be realized via the process of spontaneous parametric down-conversion (SPDC) in a nonlinear crystal (NLC) (for a review, see \cite{boyd,shySPDC}).

SPDC is the well-known nonlinear optical process, in which a classical pump laser beam at higher frequency is incident onto an optical nonlinear material which under the so-called phase-matching (PM) conditions, i.e., energy-momentum conservation, a twin-photons beam, i.e., an entangled pair of signal-idler photons at lower frequencies, is generated out of quantum vacuum \cite{shySPDC,boyd}. 
In contrast to the other sources such as superconductors or electromechanical systems, SPDC sources are more inexpensive, user-friendly, and work at room-temperature, need no cooling, and finally could be on-chip using integrated quantum photonic techniques to be commercialized, and that is why the people are interested in.
During these three decades the SPDC in NLCs has been a variety range of applications in quantum metrology, quantum communication, quantum computing, quantum information, and quantum thermodynamics (for more details and a review, see Ref.~\cite{quantummetrologybook}). 
Among these one can remark, for example, EPR realization \cite{epr}, the entanglement generation \cite{polarizationentanglement1995,Forbes}, quantum state teleportation \cite{teleport2001shih}, 
quantum ellipsometry \cite{quantumellipsometryexperimentm,quantumellipsometry2018}, quantum illumination \cite{qilluminationPRL,qilluminationScience,kadirradar,quantumilluminationbalaji1,electricFieldCorrelations,quantumradarWavefrontLanzagorta,quantumradarhelmy,quantumradarbalaji2}, quantum spectroscopy \cite{spectroscopy3,spectroscopy4,spectroscopy5,spectroscopy6}, squeezing generation \cite{squeezing1}, 
 quantum imaging \cite{imaging1,imaging2,imagingshy,imagingPadgett,ghostimagingthesis},
 quantum communication \cite{crypto1,crypto2,crypto3,crypto4,crypto5,crypto6,crypto8,crypto9,zbindencrypto,villoresi1}, nonlocal realism tests \cite{epr,nonlocal1,nonlocal2,nonlocal3,nonlocal4,nonlocal5,aliPPKTP,altepeter2005,jensen2013}.

Quantum state tomography (QST) is the most important approach for characterization of quantum state of system that leads to reconstruction of the density matrices operator of the system (DMOS), which includes all information of system such as probabilities and coherences. Also, the evolution of the density matrices enable us to evaluate the environmental effects on the quantum state of the system, and decoherence effects. 
It has been shown theoretically and experimentally \cite{kwiatstate,kwiatstate2,kwiatsource} that using the theory of the measurement of qubits, which includes a tomographic reconstruction of the DMOS due to the linear set of 16 polarization-measurement, together with the numerical optimization method, the so-called \textit{maximum-likelihood-technique} (MLT), one can obtain and reconstruct the non-negative definite density operator for a prepared entangled state or any other prepared quantum state in the polarization basis. By having the density operator, one can calculate all quantum coherence properties or entanglement criteria. Although there are many references which addressed two-qubit entropy measures \cite{moyaCessaEntropy2019,servatkhah2017Entropy,friisEntropy2017,shinEntropy2019}, but, very recently, it has been theoretically developed and experimentally demonstrated a new method, resource- and computationally efficient, for quantum state tomography in Fock basis via Wigner function reconstruction and semidefinite programming \cite{tomography2020}. They have shown that obtained density operator from this method is robust against the noise of measurement and relies on no approximate state displacements, and requires all physical properties. 
Furthermore, using the introduced method of Ref.~\cite{kwiatstate}, a new novel and scalable method have been developed and demonstrated \cite{tomographychip2018} for Scalable on-chip QST, which is based on expanding a multi-photon state to larger dimensionality. It leads to scale linearly with the number of qubits and provides a tomographically complete set of data with no reconfigurability.

Here in this paper, based on the above-mentioned investigations, we are motivated to try to prepare a polarization-entangled state in order to experimentally measure the entropy and criteria associated with the entanglement of the prepared quantum state.
By pumping a thin pair of type-I BBO NLCs with optical axis (OA) orthogonal to each other at wavelength 405nm and $ 45^\circ $-polarization, by tuning the optical compensator (C) to require the PM conditions, we generate approximately the maximally-entangled Bell state, a signal-idler biphoton wavepacket at wavelength 810nm, with high-brightness and degree of entanglement which shows a strong violation from the prediction of classical physics or HVTs. 
Also, by measuring the coincidence count (CC) rate of the SPDC process in a thin single-BBO NLC, we obtain the quantum efficiency of single-photon detectors (SPDs) as $ (26.7\pm 4.2) \%  $ and $ (24.4\pm 2.4) \% $, which are in good agreement to their manufacturer company. Also, we verify the linear dependency of the CC rate vs. pump power of input CW-laser, which may lead to obtaining the effective second-order susceptibility crystal by knowing other properties of NLC.
Finally, via the QST and using the MLT, we obtain the physical density matrices. Therefore, by having the DMOS we calculate the Concurrence, entanglement of formation, tangle, linear entropy, Von-Neumann entropy, Renyi 2-entropy, and logarithmic negativity, as the quantum coherence criteria of our prepared quantum state of two entangled photonic qubits. This high-brightness and low entangled photons rate could be used for short-range quantum measurements.

The paper is organized as follows. In Sec.~(\ref{sec2}), we present the theory of generation of the entangled state by the SPDC process in NLC. In Sec.~(\ref{sec3}), the experimental setup and Bell test results are shown. Reconstruction of density operator and QST of our prepared entangled state is also presented in Sec.~(\ref{sec3}). Moreover, in this section after obtaining the density operator we calculate the entanglement measures as well as entanglement entropy measures. Finally, we summarize our conclusion in Sec.~(\ref{summary}).

\section{Theory} \label{sec2}

\subsection{Hamiltonian and the density operator}
The Hamiltonian describing the type-I process of the SPDC which leads to producing a two-mode squeezed and polarization-entangled signal-idler photons can be written as (for more details see Refs.~(\cite{shySPDC,aliDNA,acinQKD2018})
\begin{eqnarray}\label{hamiltonianspdc}
&& \hat H= ig (\hat a^{\dag (s)}_H \hat a^{\dag (i)}_H + \hat a^{\dag (s)}_V \hat a^{\dag (i)}_V ) +\rm h.c. , 
\end{eqnarray}
where $ g $ is a coupling frequency and depends on the second order nonlinearity of the crystal \cite{acinQKD2018}, $ \hat a_{H(V)}^{(s(i))}$ is the bosonic annihilation operator of the spatial mode $ s(i) $ with $ H $ ($ V $)  polarization.
Using the su(1, 1) algebra operators which obey $ [\hat L_-, \hat L_+] = 2 \hat L_z $ and $ [\hat L_z, \hat L_\pm] = \pm L_\pm $ with $ \hat L_+=\hat a^{\dag (s)}_H \hat a^{\dag (i)}_H + \hat a^{\dag (s)}_V \hat a^{\dag (i)}_V $ and $ \hat L_z= \hat N +\frac{1}{2} \hat 1  $ where $ \hat N=\hat n_H^{(s)} + \hat n_V^{(s)} +\hat n_H^{(i)} +\hat n_V^{(i)} $, one can rewrite the Hamiltonian, thus, it is straightforward to verify that the two-mode squeezed and entangled state produced via the SPDC can be written as
\begin{eqnarray} \label{stateSPDC1}
&& \vert \Psi \rangle =e^{-i\hat H t} \vert 0 \rangle= e^{\tau( \hat L_+ - \hat L_-)} \vert 0 \rangle= (1-p^2) e^{p \hat L_+} \vert 0 \rangle ,
\end{eqnarray}
where $ \vert 0 \rangle $ denotes the vacuum state of all modes, $ q=\tanh^2\tau $ being the effective parameter of the SPDC process and dimensionless parameter $ \tau=g t $ can be assumed to be real and positive. 
Experimentally, the controllable parameter $ q $ is kept small below $ 10^{-2} $, i.e., $ \tau$  below $10^{-2} $. 
The entangled-state of SPDC, Eq.~(\ref{stateSPDC1}) can be interpreted as an incoherent mixture of different Fock states. Therefore,
the normalized density operator of the type-I entangled state due to the SPDC process can be written as \cite{acinQKD2018}
\begin{eqnarray} \label{state1SPDC}
&& \hat \rho_{\rm SPDC}^{\rm (En,I)}=(1-q^2) \sum_{n=0} (n+1) q^n \vert \psi_n \rangle \langle \psi_n \vert,
\end{eqnarray}
where $ \vert \psi_n \rangle =\frac{1}{n! \sqrt{n+1}}\hat L_+^n \vert 0 \rangle   $ is the pure state when n photon-pair excitations occur during the process of down-conversion. For simplicity, one can rewrite it in non-normalized form as follows
\begin{eqnarray} \label{statefinalSPDC}
&& \hat \rho_{\rm SPDC}^{\rm En,I}= \sum_{n=0} \frac{n+1}{2^n} p^n \vert \psi_n \rangle \langle \psi_n \vert \nonumber \\
&& \qquad  \quad = \vert 0 \rangle \langle 0 \vert +p \vert \psi_1 \rangle \langle \psi_1 \vert + O(p^2),
\end{eqnarray}
such that $ \rm Tr \rho_{\rm SPDC}^{\rm En,I}=(1-2p)^{-2} $, and  $ p=2q $ is the contribution of the desired type-I entangled Bell-state (EPR-state) due to the SPDC as 
\begin{eqnarray} \label{stateSPDCentanglement}
&& \vert \psi_1 \rangle:= \vert \psi_I \rangle= \frac{1}{\sqrt{2}} (\vert 1_H \rangle_s \vert 1_H \rangle_i + \vert 1_V \rangle_s \vert 1_V \rangle_i)=  \nonumber \\
&& \qquad \qquad = \frac{1}{\sqrt{2}} (\vert H \rangle_s \vert H \rangle_i + \vert V \rangle_s \vert V \rangle_i).
\end{eqnarray}
Note that the above state is a fully non-separable state as the basic definition of the entangled state. 
This non-factorisable state means that the state of one signal-photon cannot be specified without making reference to the other idler-photon and vise versa.
On the other hands, although a large value of $ p $ increases the production rate of the target maximally entangled two-photon states due to the SPDC, $ \vert \psi_1 \rangle $, it also increases the relative contribution of multi-photon states, $   \vert \psi_{n>1} \rangle $, due to the SPDC process. Our goals in this work are preparation of the entangled state near to the maximally entangled-state $ \vert \psi_I \rangle $ and its characterization.

Let us compare the state of the entangled photons via the SPDC and the non-normalized density operator of a single-photon (SP) source which can be written as \cite{acinQKD2018}
\begin{eqnarray} \label{statesinglephoton}
&& \hat \rho_{\rm SP}= \sum_{n=0} \bar p^{n-1} \vert n \rangle \langle n \vert= \vert 1 \rangle \langle 1 \vert + \bar p \vert 2 \rangle \langle 2 \vert + O(\bar p^2) , 
\end{eqnarray}
where $ \rm Tr \hat \rho_{SP}=(1-\bar p)^{-1} $. Note that the desired SP is produced at the zeroth order in $ \bar p $, which in the current experiment is in order of  $ 10^{-4} $, while in the SPDC process the entangle two-photon pair occurs at the first order in $ p $ in Eq.~(\ref{statefinalSPDC}).

\subsection{probability of polarization measurements}
As is seen in Eq.(\ref{stateSPDCentanglement}), the polarization of signal and idler photons can be both vertical or both horizontal with equal probability at each time. 
We can instead measure the polarization with polarizers rotated by an angle $ \alpha $ at rotated basis $ (V_\alpha,H_\alpha)^T=\bar {\bar R}(\alpha) (V,H)^T $ where $ R(\alpha) $ being rotation matrices with $ R_{11}=R_{22}=\cos \alpha $ and $ R_{12}=-R_{21}=\sin \alpha $. In this basis, the state is $  \frac{1}{\sqrt{2}} (\vert H_\alpha \rangle_s \vert H_\alpha \rangle_i + \vert V_\alpha \rangle_s \vert V_\alpha \rangle_i) $. 
It is clear if we measure in this new basis, we obtain the same results as in the old basis, half the time both are $ \vert V_\alpha \rangle $ and half another one both are in $ \vert H_\alpha \rangle  $. 
Based on the EPR interpretation of entangled-state, one can measure the signal-polarization and infer with certainty the idler-polarization. 
It should be noted that there is uncertainty between polarizations in different bases. For example, all information of a photon's polarization in the bases $ (V_{0^\circ} , H_{0^\circ})  $ implies complete uncertainty of its polarization in the bases $ (V_{45^\circ}, H_{45^\circ})  $.

Our BBO NLC is cut for type-I PM, in which the signal and idler photons emerge with the same
polarization but orthogonal to the pump photon. Note that each NLC can only support down-conversion of one pump polarization while the other polarization passes through the crystal unchanged. 
To generation of polarization-entanglement, one should use two BBO NLCs with orthogonal OA, i.e., one rotated $ 90^\circ $ from the other, so that either pump polarization can down-convert as $ \vert V_p \rangle \to \vert H_s \rangle  \vert H_i \rangle$ and $ \vert H_p \rangle \to e^{i \Delta} \vert V_s \rangle  \vert V_i \rangle$, where $ \Delta $ is a phase because of dispersion and birefringence effects in the BBO NLCs. 

Generally, to create an entangled state, we linearly polarize the pump beam at an angle $ \theta_p $ from the vertical and then shift the phase of one polarization component through passing a compensator, for example a birefringent plate, by $ \phi_c $ which leads to the pump state at crystal entrance as $ \vert \psi_{p} \rangle= \cos\theta_p \vert V_p \rangle + e^{i \phi_c} \sin \theta_p \vert H_p \rangle $. Thus, the down-converted photons state is given by
\begin{eqnarray} \label{statedown-convert}
&&  \vert \psi_{\rm DC} \rangle= \cos\theta_p \vert H_s H_i \rangle + e^{i \phi} \sin \theta_p \vert V_s V_i \rangle , 
\end{eqnarray}
where $ \phi=\phi_c +\Delta $ being the total phase difference of the two polarization components. In the experiment to generate the maximally entangled-state, one must try to generate the equal probabilities for H and V polarization, i.e., $ \theta_p=45^\circ $, means that one should illuminate the NLCs with $ 45^\circ $-polarized pump.

For measuring the polarization of down-converted signal0idler photons, we place the polarizers rotated at angles $ \alpha $ and $ b\beta $ in the signal and idler paths, respectively. Note that in our experiment the combination of a half-wave plate (HWP) at angled $ \frac{\theta}{2} $together a polarizing beam splitter (PBS) or a Rochon beam splitter plays the role of a polarizer P at angle $ \theta $. For any pair of polarizer angles there are four possible outcomes $ o_\alpha {o'}_\beta $ (with $ o,o'=H,V $) which indicate the measurement outcome one photon in polarization state $ o_\alpha $ and the other one is in $  {o'}_\beta$. Therefore, all possible probabilities of coincidence detection of a signal-idler photon pair produced in SPDC state (\ref{statedown-convert}),
 $ P_{oo'}(\alpha,\beta)= \vert  \langle o_{\alpha}^{(s)} {o'}_{\beta}^{(i)} \vert \psi_{\rm DC} \rangle  \vert^2 $ with $ o,o'=H,V $ can be easily obtained. For example, 
\begin{eqnarray} \label{pvv}
&& P_{VV}(\alpha,\beta)=\vert \sin \alpha \sin \beta \cos \theta_p + e^{i \phi} \cos \alpha \cos \beta \sin \theta_p \vert^2  , \nonumber \\ 
&& \qquad \qquad = \sin^2 \alpha \sin^2 \beta \cos^2 \theta_p  + \cos^2 \alpha \cos^2 \beta \sin^2 \theta_p \nonumber \\
&& \qquad\qquad + \frac{1}{4} \sin^2 \alpha \sin^2 \beta \sin^2 \theta_p \cos \phi 
\end{eqnarray}
where for $ \phi=0 $ and $ \theta_p=\pi/4 $ it becomes $  P_{VV}(\alpha,\beta)=\frac{1}{2} \cos^2(\beta-\alpha) $, which only depends on the relative angle $ \beta-\alpha $.
The last term proportional to $ \cos \phi $ shows the interference between the HH and VV pasts in the quantum state of light which depends on $ \Delta $, pump, signal and idler photon wavelengths, and angle, as well as crystal properties. 
Because of the finite linewidth of the input laser and also since we collect photons over a finite solid angle and wavelength range of $ \phi $. 
Therefore, we replace $ \cos \phi $ by its average $ \langle \cos \phi \rangle=\cos \phi_m $.

Let us illustrate the effect of quantum measurement on the entangled EPR-state (\ref{stateSPDCentanglement}) \cite{nonlocal4}. 
Quantum physics allows measurement of one photon to influence the state of another photon instantly. 
If the signal photon is measured at polarization angle $ \alpha $, its result will be $ H_\alpha $ or $ V_\alpha $ that each occurring half the time with the same probability. Thus, the state has collapsed at the moment of measurement to either $  \vert H_\alpha^{(s)} H_\alpha^{(i)} \rangle $ or $  \vert V_\alpha^{(s)} V_\alpha^{(i)} \rangle $.
But, it cannot determine the state of the idler; since it is the random outcome of the measurement on the signal photon which decides whether the idler to be found as $ \vert V_\alpha^{(i)} \rangle  $ or $ \vert H_\alpha^{(i)} \rangle  $. Note that the choice of polarization angle $ \alpha $ affects on the state of the idler photon such that gives it a definite polarization in the $ \vert V_\alpha^{(i)} \rangle  $ or $ \vert H_\alpha^{(i)} \rangle  $, which it did not have before the measurement.
This process describes the nonlocality, i.e., the quantum state of a subsystem changes instantly although the subsystems or particles are separated by a large distance. 
This instantaneous action at a distance leads to paradoxes about sending messages to earlier times, and is forbidden by theory of special relativity. 
Fortunately, in this case the quantum randomness prevents any paradoxes. The measurement on the signal photon cannot be observed alone in measurements on the state of the idler photon.
After measuring the signal photon, the idler photon is equally likely to be $  H_\alpha $ or $ V_\alpha  $. Therefore, a polarization measurement at any angle $ \beta $, for example, leads to find $ V_\beta $ with probability $ P_V(\beta)= \frac{1}{2} \vert \langle V_{\beta} \vert V_{\alpha} \rangle^2 + \langle V_{\beta} \vert H_{\alpha} \rangle^2 \vert = \frac{1}{2}[\cos^2(\beta-\alpha) + \sin^2(\beta-\alpha)] = \frac{1}{2} $. As is evident, this measurement gives no information about the angle $ \alpha $. Even if the signal had not been measured, we still would find its probability the same. In this interpretation, the results of quantum measurement on the EPR-entangled-state are consistent with relativistic causality. 
It should be noted this consistency is achievable by balancing: (i) the particles \textit{nonlocally} influence each other, and (ii) the nature randomness prevents from sending messages that way, which implies on the so-called \textit{spooky actions at a distance} \cite{nonlocal4} in quantum theory.

\subsection{ A local realistic HVT and Bell's inequality as the nonlocal reality}
Some physicists, the most famous one was Einstein, had been believed that there could be existed a realistic, complete and local theory to replace quantum mechanics. In the proceeding we describe such a theory, a local realistic HVT as a classical-like theory [see Refs.~(\cite{bell,aspectbell})]. 
In this HVT, each photon with a polarization angle $ \epsilon $ does not behave similar to polarization in quantum mechanics. Whenever a photon incident on a polarizer set to an angle $ \gamma $, it will always register as $ V_\gamma $ if $ \epsilon $ is closer to $ \gamma $ than to $ \gamma + \pi/2 $ such that for $ \vert \gamma - \epsilon \vert \le \pi/4 $ and $ \vert \gamma - \epsilon \vert >3\pi/4 $, $ P_V^{(\rm HVT)}(\gamma,\epsilon)=1 $, otherwise it is zero. In type-I BBO, the signal and idler have the same polarization $ \epsilon_s=\epsilon_i=\epsilon $. 
When pairs are generated, $ \epsilon $ changes in an unpredictable manner that uniformly covers the whole range of possible polarizations.
Note that $ \epsilon $ is the \textit{hidden variable}, i.e., a piece of information that is absent from quantum mechanics.

HVTs have no spooky properties of quantum mechanics and are local, it means that the measurement outcomes are determined by properties at the measurement site.
Any measurement on the signal (idler) photon is labeled by $ \epsilon_s $ and $ \alpha $ ($ \epsilon_i $ and $ \beta $). 
A realistic theory implies on that all measurable physical quantities are independent of our knowledge of them, and should have definite values. 
Furthermore, for a given $ \epsilon  $, the theory is \textit{complete} and specifies all of the possible values.
Moreover, there is no constrain on $ \epsilon $ to be random and changes deterministically such that remains to be discovered in HVTs.

To compare HVT to quantum mechanics, one should predict the coincidence probability which occurs when $ \epsilon $ is in a range such that both polarization angles $\alpha $ and $ \beta $ are close to $ \epsilon $. This coincidence probability is given by [to see general HVT approach see Ref.~(\cite{nonlocal4})]
\begin{eqnarray} \label{phvt}
&&  P_{VV}^{\rm (HVT)}(\alpha,\beta)= \frac{1}{\pi} \int_0^\pi P_{V}^{\rm (HVT)}(\alpha,\epsilon) P_{v}^{\rm (HVT)}(\beta,\epsilon) d\epsilon \nonumber \\ 
&& \qquad \qquad \qquad = \frac{1}{2} - \frac{\vert \beta - \alpha \vert}{\pi}.
\end{eqnarray}
The above coincidence probability obtained from HVT is linear vs. $ \vert  \beta - \alpha \vert $ while its quantum analogous depends on $ \cos^2(\beta - \alpha) $.
As is evident, both predictions are fairly similar, but where they disagree, quantum mechanics predicts stronger correlations or stronger anti-correlations than any HVTs.

Note that HVT introduced here is very simple, and it agrees pretty well with quantum mechanics \cite{nonlocal4}. 
Although it might be thought that some slight modification would bring our HVT into perfect agreement with quantum mechanics, in 1964, Bell showed that this is impossible \cite{bell,aspectbell}. He derived his inequality, the so-called \textit{Bell}'s inequality, and showed all HVTs obey, while quantum mechanics violates. 
Here, in the proceeding, we will use CHSH (Clauser, Horne, Shimony, and Holt) inequality \cite{chshbell} which is slightly different from the original Bell's inequality, but it is still called Bell inequality.

Bell introduced an inequality which constrains the degree of polarization correlation under measurements at different polarizer angles. Bell inequality includes two measures $ E(\alpha,\beta) $ and $ S $. The first one incorporates all possible measurement outcomes and varies from $ -1 $ (when the polarizations always agree) to $ +1 $ (when the polarizations always disagree) and is given by \cite{chshbell,bell}
\begin{eqnarray} \label{ecorrelation}
&& \!\!\!\!\!\!\!\!\!\!\! E(\alpha,\beta)= \!  P_{VV}(\alpha,\beta) \! + \! P_{HH}(\alpha,\beta) \! - \! P_{VH}(\alpha,\beta) \! - \! P_{HV}(\alpha,\beta).
\end{eqnarray}
To find the E measure, one needs to calculate the above probabilities $P_{VV}(\alpha,\beta)= N(\alpha,\beta)/N_{\rm tot}$, $ P_{VH}(\alpha,\beta)= N(\alpha,\beta_{\perp})/N_{\rm tot} $, $ P_{HV}(\alpha,\beta)= N(\alpha_{\perp},\beta)/N_{\rm tot} $ and $ P_{HH}(\alpha,\beta)= N(\alpha_{\perp},\beta_{\perp})/N_{\rm tot} $ where $ N_{\rm tot}= N(\alpha,\beta)+ N(\alpha_{\perp},\beta) + N(\alpha,\beta_{\perp}) + N(\alpha_{\perp},\beta_{\perp}) $ is the total number of pairs detected (with $ o_\perp=o + 90^\circ $).

Note that measuring $ E $ requires coincidence measurement in equal time intervals with the polarizers set four different angles. In this manner, we assume that the flux of photon pairs is the same in each interval and is not dependent on the polarizer settings. Although they create a loophole, but, are sensible and reasonable. 
Under the assumption that the polarizer settings influence the rate of down-conversion, an HVT can account for any results we observe. 
But, there is no evidence to support such a hypothesis. Nevertheless, by accepting locality and realism, an ad hoc hypothesis of this sort may be more plausible than the alternative.

The quantity $ E(\alpha,\beta) $ requires four N measurements and can be simplified as 
\begin{eqnarray} \label{ecorrelation2}
&& \!\!\! \!\!\!\! E(\alpha,\beta)= \frac{N(\alpha,\beta) +N(\alpha_{\perp},\beta_{\perp}) - N(\alpha,\beta_{\perp}) - N(\alpha_{\perp},\beta)}{N(\alpha,\beta) +N(\alpha_{\perp},\beta_{\perp}) + N(\alpha,\beta_{\perp}) + N(\alpha_{\perp},\beta)} , 
\end{eqnarray}

The second measure $ S $ is given by \cite{chshbell,bell}
\begin{eqnarray}
&& S= \vert E(a,b) - E(a,b')\vert + \vert  E(a',b) + E(a',b')\vert ~ ,
\end{eqnarray}
where $ a,b,a',b' $ are different polarizer angles, and requires sixteen coincidence count measurement. It should be noted that the measure $ S $ is interestingly theory-independent and has no clear physical meaning. Its importance originates from the fact that for any HVT and arbitrary angles it can be proved that $ \vert S \vert \le 2 $ \cite{nonlocal4,chshbell,bell}. 
Surprisingly, it can be shown that quantum mechanics can violate this inequality by considering non-maximally entangled-state (\ref{statedown-convert}) or can maximize it for the maximally entangled-state (\ref{stateSPDCentanglement}), which is equivalent to the visibility of coincidence probability $ \mathcal{V}>0.71 $ in both H-V and D-A basis (here, visibility is defined as $ \mathcal{V}= \frac{\vert P_{HH}-P_{VV} \vert }{P_{HH}+P_{VV} }$ or $ \frac{\vert P_{DD}-P_{AA} \vert }{P_{DD}+P_{AA} } $ ). By choosing the polarizer angles $ a=-45^\circ $, $ b=-22.5^\circ $, $ a'=0 $ (V), and $ b'=+22.5^\circ $, one can find that $ S^{\rm (QM)}=2\sqrt{2} $ for any maximally entangled EPR-state (\ref{stateSPDCentanglement}). Other quantum states give lower values of $ S_{\! max}=S^{\rm (QM)} $. Interestingly, for these angles, our simple HVT gives the maximum S-value $ S^{(\rm HVT)} =2$ (It can be proved for any general HVT).

The Bell inequality emphasizes that no local and realistic theory ( or complete theory in the EPR context) will ever agree with quantum mechanics. 
The challenge of whether nature agrees with quantum mechanics or any HVT or classical theory which obeys the Bell inequality can be verified by measuring $ S- $value of entangled state generated by a NLC or any source of entanglement which produces the entangled EPR-state like (\ref{stateSPDCentanglement}) or (\ref{statedown-convert}). 
If $ S > 2 $ ($ \mathcal{V} >0.71 $), we will have violated the Bell inequality and disproved all HVTs or classical theory. While if we find $ S< 2 $, both quantum mechanics and HVTs or classical theory are consistent with this result. 
To make sure from the reliability of the measurement of the violation, one should compute the statistical uncertainty of measured $ S $. By considering $ \sigma_{N_i}=\sqrt{N_i} $, it can be easily obtained the uncertainty as $ \sigma_S= \sqrt{\sum_{j=1}^{16} (\sigma_{N_i} \frac{\partial S}{\partial N_i})^2 }= \sqrt{\sum_{j=1}^{16} N_i (\frac{\partial S}{\partial N_i})^2 } $ where $\frac{\partial S}{\partial N_i}=\frac{dS}{dE} \frac{\partial E}{\partial N_i} $, $ \frac{\partial E}{\partial N_1}=\frac{\partial E}{\partial N_2}=2\frac{N_3 + N_4}{N_{tot}} $ and $ \frac{\partial E}{\partial N_3}=\frac{\partial E}{\partial N_4}=-2\frac{N_1 + N_2}{N_{tot}} $ (note that violation more that 3$  \sigma_S $ is reliable).

\subsection{Theoretical approach for quantum state tomography of two-entangled qubits \label{state}}

\subsubsection{Single-qubit}
There is a direct analogy between the measurement of the density matrix of any qubits and the measurement of the polarization state of a light beam \cite{kwiatstate}.
The polarization of the light can be characterized by the \textit{Stokes} parameters, which are evaluated from a set of four intensity measurements (i) the half intensity of incident light regardless of its polarization, (ii) intensity of the horizontal polarization; (iii) intensity of diagonal ($ 45^\circ $) polarization and (iv) right-circularly polarization. Note that the classical intensity is equivalent to the number of photons counted by a detector. These are given by 
\begin{subequations} \label{stoksN}
	\begin{eqnarray}
	&&\!\!\!\!\!\!\!\! n_0= \frac{N}{2}  (\langle H \vert \hat \rho \vert H \rangle + \langle V\vert \hat \rho \vert V \rangle)= \frac{N}{2} (\langle R \vert \hat \rho \vert R \rangle + \langle L \vert \hat \rho \vert L \rangle), \\
	&& \!\!\!\!\!\!\!\! n_1= \frac{N}{2}  (\langle H \vert \hat \rho \vert H \rangle) \nonumber \\ 
	&&  = \frac{N}{2}(\langle R \vert \hat \rho \vert R \rangle + \langle L \vert \hat \rho \vert L \rangle +\langle L \vert \hat \rho \vert R \rangle+ \langle R \vert \hat \rho \vert L \rangle), \\
	&&\!\!\!\!\!\!\!\! n_2= \frac{N}{2}  (\langle A \vert \hat \rho \vert A \rangle) \nonumber  \\
	&& = \frac{N}{2}(\langle R \vert \hat \rho \vert R \rangle + \langle L \vert \hat \rho \vert L \rangle -i \langle L \vert \hat \rho \vert R \rangle + i \langle R \vert \hat \rho \vert L \rangle), \\
	&&\!\!\!\!\!\!\!\! n_3= \frac{N}{2}  (\langle R \vert \hat \rho \vert R \rangle),
	\end{eqnarray}
\end{subequations}
where $ \vert H \rangle= (1,0)^T $, $ \vert V \rangle= (0,1)^T $, $ \vert A \rangle=(\vert H \rangle-\vert V \rangle)/\sqrt{2}=e^{i\pi/4} (\vert R \rangle +i \vert L \rangle)/\sqrt{2} $, $ \vert R \rangle= (\vert H \rangle- i\vert V \rangle)/\sqrt{2} $ are the linear horizontal, vertical, linear anti-diagonal ($ -45^\circ $), and right-circular polarization state and $ \hat \rho $ is the 2D  density matrix for the polarization degrees  (or for a two-level quantum system) and N is a constant dependent on the detector efficiency and initial light intensity. 
The Stokes parameters vector is defined as $ \vec {\mathcal{S}}= (\mathcal{S}_0,\mathcal{S}_1,\mathcal{S}_2,\mathcal{S}_3)^T $ where $ \mathcal{S}_0=2n_0 $, $ \mathcal{S}_1=2(n_1-n_0) $, $ \mathcal{S}_2=2(n_2-n_0) $ and $ \mathcal{S}_3=2(n_3-n_0) $. Thus, one can now find the density matrix with respect to the Stokes parameters as 
\begin{eqnarray}
&&  \hat \rho=\frac{1}{2 \mathcal{S}_0}  \vec {\mathcal{S}}.\vec {\hat \sigma}, 
\end{eqnarray}
where $ \sigma_0= \rm \hat I $ is the identity matrix, and $ \sigma_j $ ($ j=1,2,3 $) are the Pauli spin operators. In this manner, the tomographic measurement of the density matrix of a single qubit is equivalent to measuring the Stokes parameters.

\subsubsection{Two-qubits}
We are going to generalize the Stokes approach to measure the quantum state of two entangled qubits. Before this, let us illustrate the differences between single photon, i.e., one qubit, and two entangled qubits. Indeed, a single-photon density matrix can be mapped to the purely classical concept of the coherency matrix, while for a two qubits, one should consider the possibility of nonclassical correlations, for example, quantum entanglement.

For the two-qubit state, one can write the density operator as the tensor product of each qubit as follows \cite{kwiatstate}
\begin{eqnarray} \label{density1}
&& \hat \rho_{AB}=\frac{1}{2^2} \sum_{i,j=0}^3 r_{ij} ~ \hat \sigma_i^{(A)} \otimes \hat \sigma_j^{(B)} ,
\end{eqnarray}
where $ 4^2 $ parameters $ r_{ij} $ are real with normalization condition $ r_{00}=1 $ and also can be specified by $ 4^2-1 $ independent real parameters. 
As shown, the state of a single-qubit can be determined by a set of 4 projection operators ($ {\hat \mu_j} $) $ \hat \mu_0= \vert H \rangle \langle H \vert + \vert V \rangle \langle V \vert $, $ \hat \mu_1= \vert H \rangle \langle H \vert $, $ \hat \mu_2= \vert A \rangle \langle A \vert $ and $ \hat \mu_3= \vert R \rangle \langle R \vert $. However, it should be noted that this set of Stokes projection is at all unique, for example, one can use the set $ \hat \mu_j: $, $ \hat \mu_0= \vert H \rangle \langle H \vert $, $ \hat \mu_1= \vert V \rangle \langle V \vert $, $ \hat \mu_2= \vert D \rangle \langle D \vert $ and $ \hat \mu_3= \vert R \rangle \langle R \vert $. 
Therefore, the state of two-qubits can be determined by the set of 16 measurement projector operators $ \hat \mu= \hat \mu_i \otimes \hat \mu_j $ with $ i,j=0,1,2,3 $. Note that the outcome of a measurement $ n $ ($ n_{ij} $) is $ n=\mathcal{N} {\rm tr}(\hat \rho \hat \mu) $ where $ \mathcal{N} $ is a constant of proportionality which can be obtained from the experimental data. It worths to mention that the single-qubit measurement operators are linear combinations of the Pauli operator as $ \hat \mu_i= \sum_{j=0}^3  \Upsilon_{ij} \hat \sigma_j$ where $ \Upsilon $ and its left inverse are given by \cite{kwiatstate}
\begin{eqnarray} \label{gammamatrix}
&& \Upsilon = \left( \begin{matrix}
{1} & {0} & {0} & {0}   \\
{\frac{1}{2}} & {\frac{1}{2}} & {0} & {0}  \\
{\frac{1}{2}} & {0} & {\frac{1}{2}} & {0} \\
{\frac{1}{2}} & {0} & {0} & {\frac{1}{2}}  \\
\end{matrix} \right), \qquad    \Upsilon^{-1} = \left( \begin{matrix}
{1} & {0} & {0} & {0}   \\
{-1} & {2} & {0} & {0}  \\
{-1} & {0} & {2} & {0} \\
{-1} & {0} & {0} & {2}  \\
\end{matrix} \right), 
\end{eqnarray}
where $ \Upsilon^{-1}_{ik} \Upsilon_{kj}=\delta_{ij} $. In this manner, one can find the outcome of measurement as \cite{kwiatstate}
\begin{eqnarray} \label{r}
n \equiv n_{ij}=\mathcal{N} \sum_{l,m=0}^3 \Upsilon_{il} \Upsilon_{jm} r_{ij} .
\end{eqnarray}
Now, analogous to the single photon Stokes parameters, one can introduce the 2-photon Stokes parameter as \cite{kwiatstate}
\begin{eqnarray}\label{stoks2}
&& \mathcal{S}_{ij}\equiv \mathcal{N} r_{ij}= \sum_{l,m,s,n=0}^3 n_{ij} ({\Upsilon^{-1}})_{lm} ({\Upsilon^{-1}})_{sn},
\end{eqnarray}
with normalization condition  $ \mathcal{S}_{00}= \mathcal{N} $. Thus, the 2-qubit density matrix can be rewritten in terms of the Stokes parameters as follows
\begin{eqnarray} \label{densitystokse}
&& \hat \rho_{AB}=\frac{1}{2^2} \sum_{i,j=0}^3 \frac{\mathcal{S}_{ij}}{\mathcal{N}} ~ \hat \sigma_i^{(A)} \otimes \hat \sigma_j^{(B)} .
\end{eqnarray}

\subsubsection{Tomographic reconstruction of entangled polarization-state}
To experimentally realize and measure the different set of projection on the generated entangled photon pairs for any polarization state, one can use a combination of a polarizer which transmits only vertically polarized light (for example a PBS), a quarterwave plate (QWP), and a half-wave plate (HWP) which are placed in the beam in front of each single-photon detector (SPD). 

Using the Jones matrix one can write the projection state as \cite{kwiatstate}
\begin{eqnarray} \label{projectionstate1}
&& \vert \psi_\nu \rangle \equiv \vert \psi_{\rm proj}^{(AB)}(h_1,q_1;h_2,q_2) \rangle=  \vert \psi_{\rm proj}^{(A)}(h_1,q_1) \rangle \otimes \vert \psi_{\rm proj}^{(B)}(h_2,q_2) \rangle, \nonumber  \\ 
\end{eqnarray}
which is equivalent to the projection measurement operator as $ \hat \mu_{\nu}= \vert \psi_\nu \rangle  \langle \psi_\nu \vert $ with $ \nu=1,2,...,16 $. The projected state of each qubit, $ \vert \psi_{\rm proj}^{(1)}(h,q) \rangle $, is given by
\begin{eqnarray} \label{projectionstate2}
&& \vert \psi_{\rm proj}^{(1)}(h,q) \rangle \! = \! U_{\rm QWP}(q) U_{\rm HWP}(h) \vert V \rangle \nonumber  \\
&& \qquad \qquad \quad  = a(h,q) \vert H \rangle +  b(h,q)  \vert V \rangle ,
\end{eqnarray}
with 
\begin{eqnarray} \label{a,b}
&& a(h,q)=\frac{1}{\sqrt{2}} [\sin 2h - i \sin (2(h-q))] , \\
&& b(h,q)=-\frac{1}{\sqrt{2}} [\cos 2h + i \cos (2(h-q))] ,
\end{eqnarray}
where $ h,q $ are, respectively, the angle of the fast-axis of HWP and QWP concerning the vertical axis-polarization [measured from the vertical. Note that The angles of the fast axes can be set arbitrarily, so allowing the V-projection state fixed by the polarizer to be rotated into any arbitrary polarization state]. 
Consequently, the normalized coincidence count (CC) rate ($ s_\nu $) in the experimental measurement is given by \cite{kwiatstate}
\begin{eqnarray} \label{n_nu}
&& s_\nu=\frac{n_\nu}{\mathcal{N}}=\frac{ N_{cc}^{(\nu)}}{\mathcal{N}}= \langle \psi_\nu \vert \hat \rho \vert \psi_\nu \rangle. 
\end{eqnarray}
where $ \mathcal{N} $ is a constant parameter dependent on the photon flux and quantum efficiencies of SPDs.

For simplicity in the experiment, let us try to convert the 4$ \times $4 density matrix into a 16-dimensional column vector. Using the generators of the Lie algebra SU(2)$ \otimes $SU(2), one can introduce a set of 16 linearly independent $ 4 \times 4 $ matrices $ \hat \Gamma_{\nu}= \frac{1}{2} \hat \sigma_i \otimes \hat \sigma_j $ ($ \nu=1,2,...,16 $) [two-qubit Pauli matrices] with property $ {\rm tr}(\hat \Gamma_{\nu} . \hat \Gamma_{\mu} ) = \delta _{\nu \mu}$. Thus, it is easy to show that 
\begin{eqnarray} \label{density2}
&& \hat \rho = \sum_{\nu=1}^{16} r_\nu \hat \Gamma_{\nu},     \qquad   r_\nu= {\rm tr} (\hat \Gamma_\mu . \hat \rho), 
\end{eqnarray}
 where $ r_\nu $ is the nth element of a 16-element column vector of modified density operator. After some manipulations, one finds the relation between the CC rate in the experiment ($ n_\nu $) and the elements of the vector $ r_\nu $ as $ n_\nu= \mathcal{N} \sum_{\mu=1}^{16} B_{\nu \mu} r_\mu$ where $ B_{\nu \mu}=\langle \psi_\nu \vert \hat \Gamma_\mu \vert \psi_\nu \rangle $ is a $ 16 \times 16 $ matrix. Note that for the completeness of the set of tomographic states $ {\vert \psi_\nu} \rangle $, if $ B $ is nonsingular, thus, $ r_\nu= \mathcal{N}^{-1} \sum_{\mu=1}^{16} (B^{-1})_{\nu \mu} n_\mu $. Finally, the tomographic reconstruction of the density matrix can be rewritten as
 \begin{eqnarray} \label{densityfinal}
 && \hat \rho= \frac{1}{\mathcal{N}} \sum_{\nu=1}^{16} \hat M_\nu n_\nu= \sum_{\nu=1}^{16} \hat M_\nu s_\nu 
 \end{eqnarray} 
 where 
 \begin{eqnarray} \label{matrixM}
 && \hat M_{\nu} = \sum_{\nu=1}^{16} (B^{-1})_{\nu \mu} \hat \Gamma_\mu, \\
 && \mathcal{N}=  \sum_{\nu=1}^{16} n_\nu {\rm tr } (\hat M_\nu), 
 \end{eqnarray}
in which we have used the identity relation $ \sum_{\nu=1}^{16} {\rm tr}(\hat M_\nu) \vert \psi_\nu \rangle \langle \psi_\nu \vert= \hat I$ and $ \sum_{\nu=1}^{16} \hat M_\nu= \hat I $. Note that for the introduced set of tomographic, one can show that $ {\rm tr}(\hat M_\nu)=1 $ for $ \nu=1,2,3,4 $ and otherwise is zero. Therefore, 
\begin{eqnarray} \label{normalizationN}
&& \mathcal{N}= \sum_{j=1}^{4} n_j= \langle HH \vert \hat \rho \vert HH \rangle + \langle VV \vert \hat \rho \vert VV \rangle  \nonumber  \\
&& \qquad \qquad \qquad+ \langle HV \vert \hat \rho \vert HV \rangle +\langle VH \vert \hat \rho \vert VH \rangle .
\end{eqnarray}

Note that the reconstructed density matrix for all \textit{physical} quantum states must satisfy all conditions: 

(i) Hermiticity $ \hat \rho^\dag = \hat \rho $
	
(ii) non-negative semidefinite eigenvalue in the interval [0,1] ($ \lambda_\rho \le 1$)
	
(iii) $ {\rm tr}(\hat \rho)=1 $ and also the important condition
	
(iv) $0 \le {\rm tr}(\hat \rho^2) \le 1 $.

Here, it should be noted that \cite{kwiatstate} usually in experiments due to the experimental inaccuracies and statistical fluctuations of CC rates measurement, which means that in real experiment, recorded count rates are unfortunately different from those theoretically expected values in Eq.~(\ref{n_nu}). That is reason this tomographic approach provides an unphysical density matrix that usually satisfies no one of the last three constraints.
That is why in the next subsection, we introduce a numerical optimization method, the so-called MLT \cite{kwiatstate}, to estimate and produce the physical density matrix, which uses the reconstructed density matrix from the tomographic method as a starting point and thus satisfies all the physical constraints.

\subsubsection{Maximum-likelihood-technique}
As mentioned, the tomographic-based reconstruction of the density matrix usually violates necessary physical constrains such as positivity or $ {\rm tr}(\hat \rho^2) \le 1 $. 
To solve this problem, by exploiting the tomographic matrix, as a starting point, we employ the maximum likelihood estimation technique, as a numerical optimization method, to reproduce the physical density matrix.

To estimate the maximum likelihood density matrix, we follow the below steps \cite{kwiatstate}
\begin{itemize}
	\item 
	Generate a normal, Hermitian and nonnegative physical density matrix as a function of 16-real variables $ \hat \rho_p^{\rm MLT}(z_j) $ ($ j=1,...,16 $).
	\item 
	Introduce an optimization function, the so-called \textit{likelihood} function $ \mathcal{L}(z_j;n_j) $, which quantifies how good the physical density matrix $ \hat \rho_p^{\rm MLT}(z_j) $ is regarding the experimental measurements. This optimized likelihood function is dependent on the 16 real parameters $ z_\nu $ and of the 16 experimental measured CC rates $ n_\nu $.
	\item
	Finding an optimized set of parameters $ z_j $ using the standard numerical optimization techniques, which maximize the likelihood function, $ \mathcal{L}^{\rm (opt)}(z_j;n_j) $.
\end{itemize}

The property of non-negative definiteness can be written as $ \langle \psi \vert \hat \rho_p^{(\rm 	MLT)} \vert \psi \rangle \ge 0$ for any quantum state $ \vert \psi \rangle $. On the other hand, it can be easily proven that any normalized matrix as form $ \mathcal{F^\dag F}/{\rm tr (\mathcal{F^\dag F}) } $ is nonnegative definite and Hermitian. Using the matrix algebra, one can write the matrix $ \mathcal{F} $ as a tridiagonal matrix as follows \cite{kwiatstate}
\begin{eqnarray} \label{fmatrix}
&& \mathcal{F} = \left( \begin{matrix}
{z_1} & {0} & {0} & {0}   \\
{z_5 + i z_6} & {z_2} & {0} & {0}  \\
{z_{11}+iz_{12}} & {z_7 + i z_8} & {z_3} & {0} \\
{z_{15}+iz_{16}} & {z_{13}+iz_{14}} & {z_{9}+iz_{10}} & {z_4}  \\
\end{matrix} \right).
\end{eqnarray}
Thus, the \textit{physical} density matrix $ \rho_p^{(MLT)} $ can be written as
\begin{eqnarray} \label{physicaldensity}
&& \rho_p^{(MLT)}= \mathcal{F^\dag(\textbf{z}) F(\textbf{z})}/{\rm tr[\mathcal{F^\dag(\textbf{z}) F(\textbf{z})]} }. 
\end{eqnarray}
Also, one can find the inverse of matrix $ \mathcal{F}(\textbf{z}) $ in terms of elements of the density matrix as follows \cite{kwiatsource}
\begin{eqnarray} \label{fmatrix2}
&& \mathcal{F} = \left( \begin{matrix}
{\sqrt{\frac{\Delta}{M_{11}^{(1)}}}} & {0} & {0} & {0}   \\
 {\frac{ M_{12}^{(1)} }{\sqrt{ M_{11}^{(1)} M_{11,22}^{(2)} }}} & { \sqrt{\frac{M_{11}^{(1)}}{M_{11,22}^{(2)}}} } & {0} & {0}  \\
{ \frac{M_{12,23}^{(2)}}{\sqrt{\rho_{44}} M_{11,22}^{(2)} } } & { \frac{M_{11,23}^{(2)}}{\sqrt{\rho_{44}} M_{11,22}^{(2)} } } & { \sqrt{\frac{M_{11,22}^{(2)}}{\rho_{44}} } } & {0} \\
{ \frac{\rho_{41}}{\sqrt{\rho_{44}}} } & {\frac{\rho_{42}}{\sqrt{\rho_{44}}} } & {\frac{\rho_{43}}{\sqrt{\rho_{44}}} } & {\sqrt{\rho_{44}}}
\end{matrix} \right),
\end{eqnarray}  
where $ \Delta={\rm det}\hat \rho $, $ M_{ij}^{(1)} $ is the first minor of $ \hat \rho $ which is the determinant of the $ 3 \times 3  $ matrix formed by deleting the ith row and jth column of the reconstructed tomographic matrix $ \hat \rho $; and also $ M_{ij,kl}^{(2)} $ is the second minor, the determinant of the $ 2\times 2 $ matrix formed by deleting the ith and kth rows and jth and lth
columns of the reconstructed tomographic matrix $ \hat \rho $.

\subsubsection{The likelihood function and its numerical optimization}
As said the measurement results include a set of 16 CC rates with average values $ \bar n_\nu =\mathcal{N} \langle \psi_\nu \vert \hat \rho  \vert \psi_\nu \rangle  $. Consider that the noise on the CC rate measurements is a Gaussian probability distribution which is given by \cite{kwiatstate}
\begin{eqnarray} \label{probabilityCC}
&& \mathcal{P}(n_1, n_2,...,n_{16})= \frac{1}{N_{\rm norm}}  \prod_{\nu=1}^{16} {\rm exp} [-\frac{(n_\nu - \bar n_\nu)^2}{2\sigma_\nu}], 
\end{eqnarray}
where $ \sigma_\nu \simeq \sqrt{\bar n_\nu } $ is the standard deviation of the nth CC rate measurement, and $ N_{\rm norm} $ is the normalization constant. 
Here, in our case 
\begin{eqnarray} \label{nCClikely}
&& \bar n_\nu (z_1,z_2,...,z_{16}) = \mathcal{N}  \langle \psi_\nu \vert \hat \rho_p^{\rm (MLT)}(z_1,z_2,...,z_{16}) \vert \psi_\nu \rangle,
\end{eqnarray}
where $ \mathcal{N} $ can be obtained from equation (\ref{normalizationN}). 
Thus, the likelihood that the matrix $ \hat \rho_P^{\rm (MLT)}(z_1 ,...,z_{16}) $ can produce the measured data $ n_1,...,n_{16} $ is the probability function $ \mathcal{P}(n_1, n_2,...,n_{16}) $ which should be maximized. 
Note that instead of maximizing $ \mathcal{P}(n_1, n_2,...,n_{16}) $, which is equivalent to find the maximum of its logarithm, one can minimize its argument, and thus the likelihood function can be introduced as \cite{kwiatstate}
\begin{eqnarray} \label{likelihood}
&& \mathcal{L}(z_1,z_2,...,z_{16})= \sum_{\nu=1}^{16} \frac{[\mathcal{N} \langle \psi_\nu \vert \hat \rho_p^{\rm (MLT)}(z_1,z_2,...,z_{16}) \vert \psi_\nu \rangle - n_\nu ]^2}{2 \mathcal{N} \langle \psi_\nu \vert \hat \rho_p^{\rm (MLT)}(z_1,z_2,...,z_{16}) \vert \psi_\nu \rangle } ,  \nonumber \\
\end{eqnarray} 
which should numerically be optimized. 

For optimization, one can use the software such as MATHEMATICA oR MATLAB or Python. To do this, one needs to estimate initial values of $ z_\mu $. To this, one can use the tomographic estimate of the density matrix in the inverse relation (\ref{fmatrix2}), which allows us to determine a set of values. 
Note that since the tomographic density matrix may not be nonnegative definite, the values of the $ z_\nu$'s deduced in this manner are not necessarily real. Thus for our initial guess, we can use their real parts of the $ z_\nu$'s deduced from the tomographic density matrix.
In the experimental part, we have followed this method to reconstruct the physical density matrix and then calculate the entanglement entropies and entanglement measures.

\section{experimental results } \label{sec3}

\subsection{experimental setup}

\begin{figure} 
 	\includegraphics[width=8.5cm]{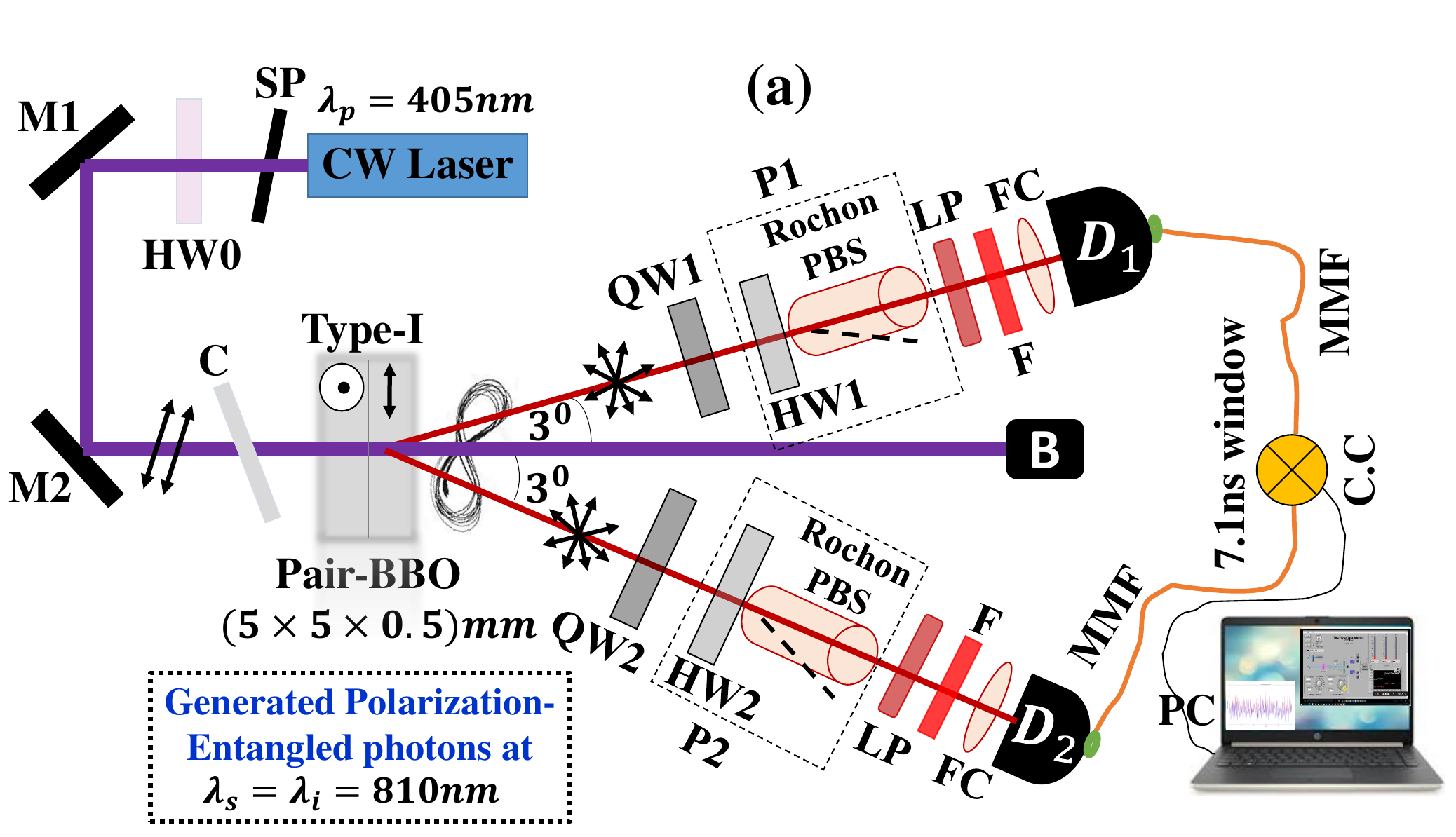}
 	\includegraphics[angle=0,width=9.5cm,origin=c]{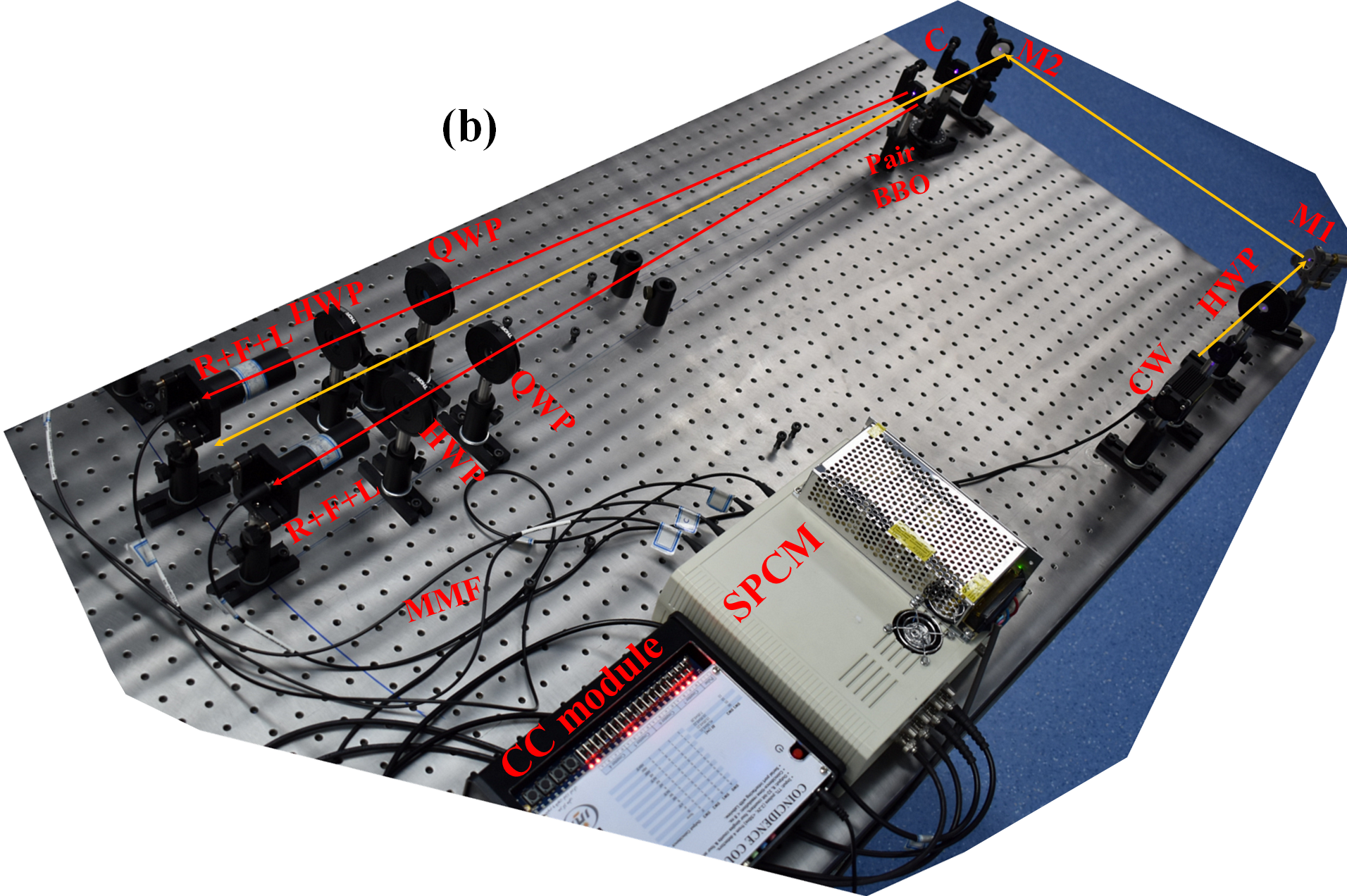}
 	\caption{(Color online) (a) Schematic illustration of the experimental arrangement for the generation of maximally entangled Bell states and quantum state tomography. (b) Experimental setup demonstration for the Bell test and quantum state tomography. By reconstructing the density matrix through the QST, we are able to calculate entanglement entropy and all entanglement measures. CW: laser diode 405nm, SP: diaphragm, HWP: half-wave plate, M1 and M2: mirror, C: compensator, QWP: quarter-wave plate, MMF: multimode fiber, R: Rochon polarizer, F:Filter; L: fiber coupler, CC module: coincidence counter and SPCM: 4-channel single-photon detector. }	
 	\label{fig1}
\end{figure}

Fig.~\ref{fig1}(a) and (b) show a schematic of our experimental setup and also the side-view of our experimental arrangement, respectively, to produce the type-I polarization-entangled photonic state. 
A $ 173{\rm mW} $ Gaussian beam of a $ 405{\rm nm} $ diode laser with diameter $ \varnothing 3 \rm mm$ after passing through a diaphragm SP1, acting as a spatial filter, passes through an HWP [Newlightphotonic: WPA03-H-810, air-spaced 0th-order waveplate at 810nm, $\varnothing $15.0mm, and OD 1" mounted] to be linearly polarized at $ +45^\circ $ (diagonal) to illuminate the paired BBO NLC. Then the pump beam is reflected from two broadband (380-420nm) high-reflective ($  R>99.0\% $) mirrors M1 and M2 [Newlightphotonic: BHR10-380-420-45; AOI 45 deg, $ \varnothing $25.4mm] and passes through a $ 5\times 5 \times 0.25$mm BBO compensator C [Newlightphotonic: BSBBO5025-405(I) matched to the 0.5mm thick Type-I SPDC crystal pumped by 405nm and with OD 1"] which is fixed on a tiltable and rotatable mount. Then, the pump beam with power around $ 80 $mW is incident vertically on the paired BBO type-I SPDC crystal [Newlightphotonic:: PABBO5050-405(I)-HA3] with each size $ 5\times5\times0.5 $mm wich cut for Type-I SPDC pumped by 405nm and with the half opening output angle of 3 degrees. Note that the two crystals optically contacted with one crystal rotated by 90 degrees about the axis normal to the incidence face. The pump beam is incident vertically on the entrance-side of the crystal such that the angle between pump wavevector and the OA is $ \rm \psi=29.3^\circ $ which yields to generate the entangled signal-idler biphoton at wavelength 810nm with output angles $ \theta_s=\theta_i\simeq 3^\circ $ in the planer case. Note that to achieve the PM condition to generate the maximally entangled Bell state, one should precisely adjust the C to exactly coincide both signal and idler wavepackets to be indistinguishable. 
In the BBO NLC through the SPDC process, a small fraction of the laser photons spontaneously decays into the signal-idler photons on opposite sides of the laser beam. It can be understood as the time-reversed process of sum-frequency generation (SFG). 
In SFG, two beams at frequencies $ \omega_1 $ and $ \omega_2 $ meet in an NLC that lacks inversion symmetry. The crystal can be  modeled as a collection of ions in anharmonic potentials. When the NLC is driven at both $ \omega_1 $ and $ \omega_2 $, the ions oscillate with several frequency components including the sum frequency $ \omega_1 + \omega_2$. Each ion radiates at this frequency. The coherent addition of light from each ion in the crystal leads to constructive interference only for specific beam directions and certain polarizations, the so-called PM condition that requires inside the crystal the wave vectors of the input beams must sum to that of the output beam.

Ultimately, the generated down-converted biphotons after passing analyzers (polarizers) P1 or P2, a HWP plus alpha-BBO-Rochon polarizer [Newlightphotonic: RPB0110 with beam deviation  about $ 15^\circ $, extinction ratio: 200,000:1, aperture 10mm, and OD 1" mounted] which act as a measurement box, reach to the single-photon counting modules (SPCMs). 
Note that the SPCMs are preceded by linear polarizers P1(P2), uncoated long-pass red filters LPFj ($ j=1,2 $) to block any scattered laser light [Newlightphotonic: LWPF1030-RG715; cutoff wavelength 715nm, $ \varnothing $25mm], and fiber coupler lenses FCj ($ j=1,2 $) [Thorlabs: F220FC-780; f = 11.07 mm, NA = 0.26 FC/PC] which are focused into the multimode fibers MMFj ($ j=1,2$) [Excelitas: SPCM-QC9FIBER-ND 100 FC], and then transferred via the couplers into the 4-channel single-photon counting module (SPCM) [Excelitas: SPCM-AQ4C; time resolution 600ps, dead-time 50ns and quantum efficiency about 28$ \% $ at 810nm]. 
Also, for QST we have placed two QWPs [Newlightphotonic: WPA03-Q-405, air-spaced 0th order quarter waveplate at 405nm] in each arm before the polarizers pack in order to measure different polarization. 

It should be note that it is necessary to use coincidence detection to separate the down-converted photons from the background of other photons reaching the detectors. Because the photons of a down-converted pair are produced at the same time, they cause coincident, that is, nearly simultaneous, firings of the SPCMs. 
Coincidences are detected by a fast logic circuit, for example, an Altera-DE2 FPGA board, with coincidence time window $ \tau=7.1 $ns and adjustable acquisition time greater than 0.1s, which are recorded by a Laptop. To make sure its validity, and also to calibration, we have compared it to the 8-channel coincidence-counter/time-tagger made by QuTools [with time resolution (bin-size) and time-stamp 81ps] such that we found their SCs and CCs were in good agreement.

The detection components of SPCMs, irises, couplers and filters are placed on a tiltable and rotatable mount which can be precisely translated in the plane perpendicular to the pump direction. 
Note that the presented arrangement allows us to adjust the SPCMs at different angles with minimal realignment and maximum CC rates. Moreover, to have the high visibility of polarization entanglement, we have placed two adjustable irises in front of each detector. The CC and SC rates are shown online in the Labview interface panel on the laptop.
All the analysis of the experimental data has been done in MATLAB software.
Finally, to decrease the background noises, we have tried to remove all the reflected and scattered light from the environment and also blocked the pump beam after crystal by a blocker B. Moreover, we control the temperature of the Lab during each experiment.
Furthermore, to keep laser power constant to avoid noise in count rates, we stabilized the diode laser by a 5V 1A driver with $ \lesssim 0.1\% $ current-fluctuation which results in $ \lesssim 0.1\% $ power-fluctuation by assumption constant load on the adapter. Thus, during our experiments, we are sure that the pump power can be assumed constant.
The distance between two detectors is about 10cm, and the distance between detectors and BBO NLC are roughly 100cm.

\subsection{calibration, tuning the Bell state and Bell inequality test}

 \begin{figure} 
 	\includegraphics[width=8.5cm,origin=c]{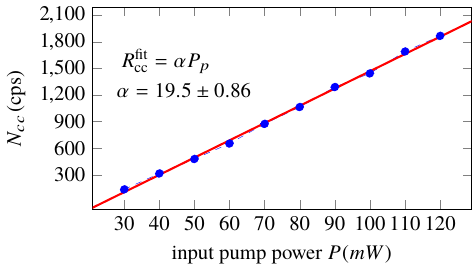}
 	\caption{(Color online) CC rate vs. input pump power. Here, we have used Malus’ law  via rotating the polarizers to change the input pump power. The dotted points show the experimental coincidences, and the red solid-line shows the fitted line with slope $ \alpha $.}	
 	\label{fig2}
 \end{figure}

After aligning the setup to have maximum CC rate, firstly, we have compared the quantum efficiency of our detectors, $ \eta_1= N_{c}/N_2 $, which $ N_c $ is the CC rate, as $ \% 26.7 \pm \% 4.2 $ ($  \% 24.4 \pm \% 2.4 $) at signal-idler wavelength 810nm to the reported efficiencies of the supplier company that have a good agreement. 

Secondly, we are motivated to verify the linear dependency of the CC rate of the generated down-converted photons via the SPDC versus pump power. 
As known, the SPDC is a quantum nonlinear process in which the zero-point quantum vacuum field stimulates an input photon to transform into a pair of correlated photons under the energy-momentum conservation (PM condition), and the probability depends quadratically on the medium second-order susceptibility $ \chi^{(2)} $.

In SPDC, unlike the SHG, the CC rate linearly and quadratically depends on the pump power and susceptibility \cite{linearcc1,linearcc2}, respectively, which are given by 
\begin{eqnarray} \label{linear1}
&& R_c= \frac{\Delta t_p}{T_p} \frac{\omega_p}{3 \pi} \frac{L^2}{A_{\rm SPDC}} \frac{P_p}{P_0}:=\alpha P_p, 
\end{eqnarray}
with 
\begin{eqnarray} \label{p0}
&& P_0=\frac{8 \epsilon_0 n_p^2 n_{\rm SPDC} c^3}{\omega_{\rm SPDC}^2 \chi_{\rm eff}^{(2) 2} }, 
\end{eqnarray}
where $ P_p $ is peak pump power at frequency $ \omega_p $, $ \Delta  t_p $ stands for the pulse width, $ T_p $ is the repetition period if the pump is not CW laser while for the CW pump, one can effectively assume $ \Delta  t_p, T_p \to \infty  $ such that $ \Delta  t_p/ T_p \to 1$. Moreover, $ n_j $ is the linear refractive indices at $ \omega_p $, $ \omega_{\rm SPDC}=\omega_p/2 $, L being the crystal length and $ A_{\rm SPDC} $ is attributed to the cross-sectional area equal to the diffraction-limited area of the pump beam. 
Fig.~(\ref{fig2}) shows clearly the linear dependency of the CC rate, $ n_C $, vs. incident pump power $ p_p $. As is seen, the behavior of the CC rate is linear with respect to the increasing input pump power. By calculating the $ \alpha $ by fitting, one can obtain and estimate the second-order susceptibility of the crystal.   \\

\begin{figure} 
	\includegraphics[width=8cm,origin=c]{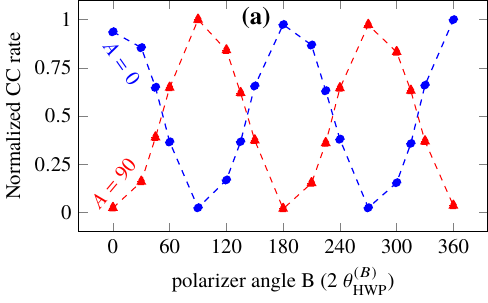}
	\includegraphics[width=8cm,origin=c]{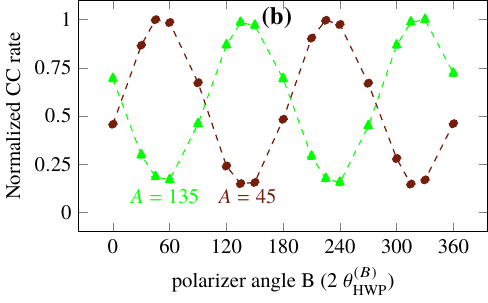}
	\caption{(Color online) Normalized CC rate of channel A (1) vs. polarizer angle (the half HWP-angle) of channel B (2). (a) A is fixed at angles $ 0^\circ $ and $ 90^\circ $. (b) A is fixed at angles $ 45^\circ $ and $ 135^\circ $. Blue and brown thick-dotted points; red and green thick-triangle are referred to polarizer angles $ A=0^\circ $, $  A=45^\circ $, $ A=90^\circ $, and $ A=135^\circ $, respectively. Note that the set of HWP together PBS acts as a polarizer such that $\theta_{\rm pol}= 2 \theta_{\rm HWP}.$}	
	\label{fig3}
\end{figure}

\begin{table} 
	\caption{ Experimental data corresponding to the Bell inequality measurement with Bell parameter $ S=2.71 \pm 0.1 $ which shows the strong violation from the HVT or any classical prediction. $ N_A $, $ N_B $ and $ N_c $ are , respectively, Singles and coincidence count rates as a function of polarizer angles in channel A and B. Here, the integration time and coincidence time-window are , respectively, $ T=0.3 \rm s $ and $ \tau=7.1 \rm ns$} 
	\begin{ruledtabular}
		\begin{tabular}{ccccccc}
			\multicolumn{2}{c}{angle \footnote{Here, angles are attributed to polarizer's angles that in our configuration are equivalent to $ \theta_{\rm HWP}/2 $. } [degree] } & \multicolumn{3}{c}{count rate [cps] } & \multicolumn{2}{c}{uncertainty [cps]}\\ 
			
			$ \theta_{\rm pol}^{A} $ & $ \theta_{\rm pol}^{B} $ & $N_A $ & $N_B $ &  $ N_c$ &  $ \Delta N_c$ & Acc. \footnote{Acc. is the accidental coincidence rate which is given by $ \tau 	N_A N_B/T. $} \\ \hline \hline

			0 & -22.5 & 3290.40 & 3938 & 27.79 & 1.53 &  0.302 \\ 
			0 & 22.5 & 3306.56 & 4082.63 & 32.15 & 1.77 &  0.314 \\ 
			0 & 67.5 & 3268.46 & 3652.60 & 6.32 & 1.47 &  0.278 \\ 
			0 & 112.5 & 3298.70 & 3569.66 & 2.05 & 1.12 &  0.274 \\ \hline

			-45 & -22.5 & 3004.56 & 3942.50 & 28.62 & 1.39 &  0.276 \\ 
			-45 & 22.5 & 2937.53 & 4038.40 & 8.38 & 1.70 &  0.276 \\ 
			-45 & 67.5 & 2906.93 & 3632.13 & 5.42 & 1.51 &  0.246 \\ 
			-45 & 112.5 & 2986.10 & 3564 & 21.21 & 1.43 &  0.248 \\ \hline

			45 & -22.5 & 2847.46 & 3897.63 & 3.27 & 1.40 &  0.258 \\ 
			45 & 22.5 & 2871.76 & 4067.20 & 26.82 & 1.49 &  0.272 \\ 
			45 & 67.5 & 2898.56 & 3652.40 & 28.35 & 1.79 &  0.247 \\ 
			45 & 112.5 & 2902.23 & 3532.23 & 7.29 & 1.69 &  0.239 \\ \hline

			90 & -22.5 & 2566.06 & 3894.56 & 6.63 & 1.71 &  0.233 \\ 
			90 & 22.5 & 2565.76 & 4059.80 & 2.4 & 1.20 &  0.243 \\ 
			90 & 67.5 & 2562.76 & 3635.46 & 25.08 & 1.76 &  0.217 \\ 
			90 & 112.5 & 2568.16 & 3529.43 & 29.75 & 1.60 &  0.211 \\ 
			
		\end{tabular}
	\end{ruledtabular}
	\label{tableBell}
\end{table}

\begin{table}
	\caption{ The results of visibility measurement. Here, the other parameters are the same as in Tab.~(\ref{tableBell}). We find $ \mathcal{V}_{\rm HV}= \%(98.50 \pm 1.35) $ and $ \mathcal{V}_{\rm DA}= \%(87.71 \pm 4.45) $ for the visibility in horizontal and diagonal bases, respectively.} 
	\begin{ruledtabular}
		\begin{tabular}{cccc} 
			$ \theta_{\rm pol}^{A} $ & $ \theta_{\rm pol}^{B} $ & $ N_c$ &  $ \Delta N_c$ \\ \hline \hline
			
			0 & 0 & 36.82 & 2.94  \\ 
			0 & 90 & 0.26 & 0.18  \\ 
			90 & 0 & 0.29 & 0.20  \\ 
			90 & 90 & 37.46 & 1.93 \\ \hline

			45 & 45 & 33.03 & 2.57  \\ 
			45 & 135 & 1.82 & 0.94  \\ 
			135 & 45 &  2.54 & 1.18 \\ 
			135 & 135 & 33.38 & 2.34 \\
			
		\end{tabular}
	\end{ruledtabular}
	\label{tableVisibility}
\end{table}

To create the maximally entangled Bell state [see Eq.~(\ref{stateSPDCentanglement})] or any other state close to it [see Eq.~(\ref{statedown-convert})], we must adjust the parameters which determine the polarization of pump laser. 
Firstly, we should close the CCs $ N_c(0,0) $ and $ N_c(90,90) $ together.
After that, one should set $ \theta_p $ by rotating the QWP about its vertical axis and also tilt it to maximize the CC $ N_c(45,45) $. Note that this optimization typically leads to collection of a few hundred photons per point, which requires an acquisition time window a few seconds. Now using the relation 
\begin{eqnarray} \label{n_cc}
&& N_c(\alpha,\beta)= N_0 P_{VV}(\alpha,\beta) + D ,
\end{eqnarray}
one can find CC $ N_C(0,0) $, $ N_C(90,90) $, $ N_C(45,45) $, $ N_C(0,90) $, $ \theta_p $ and $ \phi_m $ as follows
\begin{eqnarray}
&& D= [N_c(0,90) + N_c(0,90)]/2, \\
&& N_0=  N_C(0,0)+ N_C(90,90) - 2D , \\
&& \tan^2 \theta_p = \frac{N_c(90,90)-D}{N_c(0,0)-D} , \\
&& \cos \phi_m= \frac{1}{\sin 2\theta_p}  \left( 4 \frac{N_c(45,45)}{N_0} -1 \right). 
\end{eqnarray}
After optimization of pump polarization and the compensator angle (see the following tables which show the CC rate results in our experiment), we have found that $ D\simeq 0.275 \rm cps $, $ N_0 \simeq 73.73 \rm cps $, $ \theta_p \simeq  45.25^\circ $, and $ \phi_m \simeq 37.62^\circ $ which implies that our created state is very close to the Bell state. It should be noted that to achieve the best adjustment, one must try many times and find the above parameters to check how much we are close to the maximally entangled state.

Fig.~(\ref{fig3}) shows the behavior of the normalized CC rate when the polarizer angle of channel A(1) [or half of the HWP angle] is fixed at angles $ 0^\circ $, $90^\circ $, $ 45^\circ $ and $ 135^\circ $ for the different values of the polarizer angles of channel B(2). As is seen, our experimental points show a good agreement with the theoretical relation (\ref{n_cc}) within the quantum entanglement framework.

Table.~(\ref{tableBell}) shows a typical set of Bell measurements. As is seen, the accidental coincidence count rates [which is the average number of photons from two different down-conversion events will arrive at detectors purely by happenstance], within the coincidence interval should be omitted from the CC rate $ N_c $ in the calculations.
Note that this accidental rates background are small, nearly constant, and decreases the Bell parameter $ S $. Thus, finding $ S>2 $ never can be an artifact of the accidental background rate.
Using these experimental measurements, we obtained $ S=2.71 \pm 0.1 > 2 $, which shows a strong violation. This strong violation of Bell inequality by more than six standard deviations conclusively eliminates the HVTs, and is consistent with prediction of quantum theory.

The results of the visibility measurement can be seen in Tab.~(\ref{tableVisibility}). As is seen, the visibility $ \mathcal{V} $ of the coincidence count rate, as second-order interference, is respectively, $ \mathcal{V}_{\rm HV}= \%(98.50 \pm 1.35) $ and $ \mathcal{V}_{\rm DA}= \%(87.71 \pm 4.45) $ for the horizontal and diagonal bases. Since $ \mathcal{V}> \% 71 $, once can conclude a violation from the HVTs.

As is evidence, our  experimental results [see Tabs.~(\ref{tableBell} and (\ref{tableVisibility})) and Fig.~(\ref{fig3})] show that we could create an entangled state near to the maximally entangled Bell state $ \frac{1}{\sqrt{2}} (\vert H \rangle_s \vert H \rangle_i + e^{i\phi} \vert V \rangle_s \vert V \rangle_i) $ with the strong violation, $ S>2 $ ($ \mathcal{V} >0.71$), from the prediction of classical physics. 

Finally, let us compare our results to others. The most important parameter which might be compared is the brightness and entanglement which are characterized by the visibility, $ \mathcal{V} $, or Bell parameter, $ S $. Note that the ratio of $  S/S_max $ in our case is very large as 0.96 with respect to other similar sources. For example, in Ref.~\cite{nonlocal4}, Ref.~\cite{altepeter2005} and Ref.~\cite{jensen2013} are, respectively, 0.82, 0.93 and 0.89 which show that our quantum source is a stronger entanglement photonic source with higher brightness (visibility) which yields the higher entanglement measures.

\begin{table} 
	\caption{ Experimental measurement data for the tomographic analysis states to reconstruct the physical density matrix using the MLT. Here, we have used the notation $ \vert D \rangle = (\vert H \rangle + \vert V \rangle)/\sqrt{2} $, $ \vert L \rangle = (\vert H \rangle + i \vert V \rangle)/\sqrt{2} $ and $ \vert R \rangle = (\vert H \rangle - \vert V \rangle)/\sqrt{2} $. Note that overall phase factors cannot affect the results of projection measurements. } 
	\begin{ruledtabular}
		\begin{tabular}{ccccccc}
			\multicolumn{2}{c}{ Proj. Mode \footnote{Note that when the measurements are taken, only one wave plate angle has to be changed between measurements. Thus, for example, one can follow the arrangement of the polarization configuration $ \nu $ in the table.} } & \multicolumn{2}{c}{ WP angles} & \multicolumn{3}{c}{ count rate [cps]}\\ 
			
			$ \nu $ & $ \rm AB $ & ($ \rm h_A,q_A $) &  ($ \rm h_B,q_B $) & $ N_A $  & $ N_B $  &  $ N_c $  \\ \hline \hline

			1 & HH & (45,0) & (45,0) & 2496.41 & 3450.92 &  32.56 \\ 
			2 & HV & (45,0) & (0,0) & 2498.72 & 4006.25 &  1.01 \\ 
			3 & VV & (0,0) & (0,0) & 3139.43 & 4001.54 &  32.00 \\ 
			4 & VH & (0,0) & (45,0) & 3134.21 & 3444.14 &  1.15 \\ \hline

			5 & RH & (22.5,0) & (45,0) & 2803.92 & 3454.57 &  15.02 \\ 
			6 & RV & (22.5,0) & (0,0) & 2815.87 & 3998.55 & 20.01 \\ 		
			7 & DV & (22.5,45) & (0,0) & 2785.89 & 3988.65 &  14.63 \\ 	
			8 & DH & (22.5,45)& (45,0) & 2785.00 & 3464.92 & 21.32\\ \hline

			9 & DR & (22.5,45) & (22.5,0) & 2788.50 & 3826.91 &  21.99 \\ 
			10 & DD & (22.5,45) & (22.5,45) & 2789.82 &3662.35 &  27.48 \\ 
			11 & RD & (22.5,0) & (22.5,45) & 2809.30 & 3662.82	 &  22.82 \\ 
			12 & HD & (45,0) & (22.5,45) & 2505.73 & 3664.05 &  17.85 \\ \hline

			13 & VD & (0,0) & (22.5,45) & 3127.28 & 3668.48 &  16.20 \\ 		
			14 & VL & (0,0) & (22.5,90) & 3135.65 & 3802.62 &  18.44 \\ 
			15 & HL & (45,0) & (22.5,90) & 2498.29 & 3785.32 & 15.13 \\ 						
			16 & RL & (22.5,0) & (22.5,90) & 2817.60 & 3793.93 &  28.78\\ 
			
		\end{tabular}
	\end{ruledtabular}
	\label{tabletomography}
\end{table}

\subsection{Experimental reconstruction of the physical density matrix and calculation of entanglement measures }

In this subsection, we are going to obtain the actual physical state of our created entangled state with strong Bell violation via the QST assisted by MLT. 
As said in the theoretical subsection (\ref{state}) with details, one can extract the tomographic state of the system with using the set of 16 CC-measurement of the polarization, which can be realized by setting specific values of the half- and quarter-wave plate angles (see Tab.~(\ref{tabletomography}))). Now, by having the tomographic states measurement, as said in subsection (\ref{state}), one can obtain the physical density matrix using the numerical optimization, i.e., MLT, as follows
\begin{widetext}
	\begin{eqnarray} \label{roexperiment}
	\rho_{\rm rec} ^{ \rm (MLT)}= \left( \begin{matrix}
	\rm {HH} & \rm {HV} & \rm {VH} & \rm {VV}  \\
	{0.487} & $ 0.015+0.012i $ & $ 0.054 + 0.017i $ & $ 0.285 + 0.134i $   \\
	$ 0.015-0.012i $& {0.015} & \rm {0.005- 0.013i} & \rm {-0.024 + 0.05i}  \\
	$ 0.054- 0.017i $ &\rm {0.005+0.013i} & {0.018} & $ 0.001 - 0.025i $ \\
	$ 0.285- 0.134i $& \rm {-0.024-0.05i} & $ 0.001+ 0.025i $  & {0.480}
	\end{matrix} \right)_{4\times 4} \!\!\! \!\!\!\!\! , \nonumber \\
	\end{eqnarray}
\end{widetext}
which requires all the physical constraints. The eigenvalues of the physical density matrix are $ p_a^{(j)}=0.801,0.199,0,0 $ with j=1,2,3,4 ($ \rm tr \rho^2=0.682 < 1 $) corresponding to eigenvectors
\begin{widetext}
	\begin{eqnarray} \label{eigenvector1}
	U_{\Psi}= \left( \begin{matrix}
	\rm {\vert \psi^{(1)} \rangle } & \rm {\vert \psi^{(2)} \rangle } & \rm {\vert \psi^{(3)} \rangle } & \rm {\vert \psi^{(4)} \rangle }  \\
	{0.642+0.302i} & $ -0.587-0.272i$ & $0.249 +0.124i $ & $0.0015 $   \\
	$ -0.005+0.04i $& $-0.162+0.21i $ & $-0.374+0.373i $ & $-0.1+0.8i$ \\
	$ 0.038+0.012i $ & $ -0.156-0.243i $ & $-0.447-0.609i $ & $0.559+0.177i $ \\
	$0.702$ & $ 0.654 $ & $-0.27 $  & $ -0.0812 $
	\end{matrix} \right) . \!\!\! \!\!\!\!\! \nonumber \\
	\end{eqnarray}
\end{widetext}

\begin{figure} 
	\includegraphics[width=8cm,origin=c]{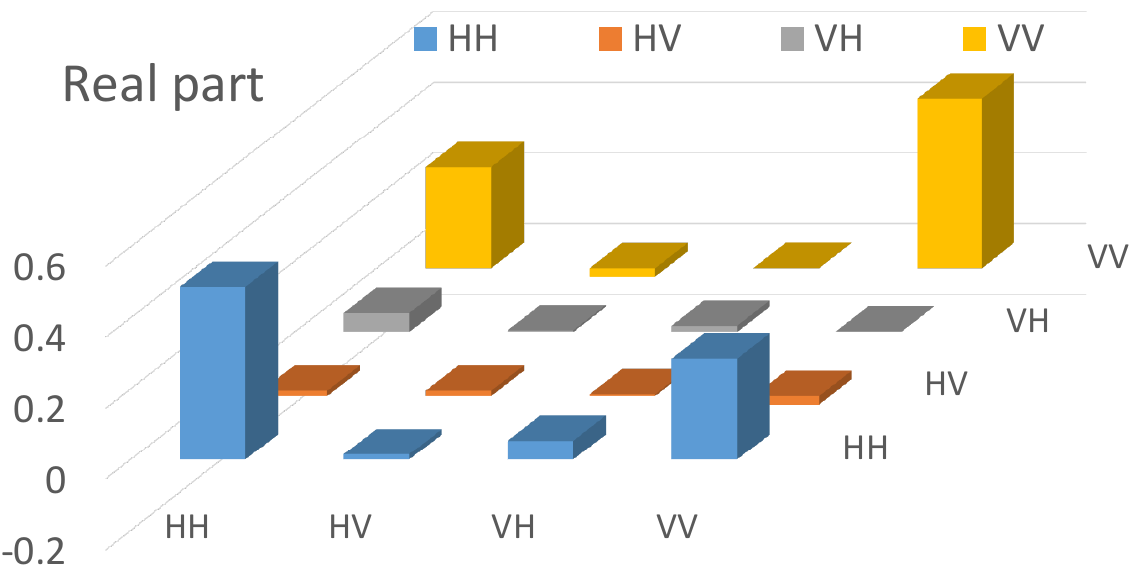}
	\includegraphics[width=8cm,origin=c]{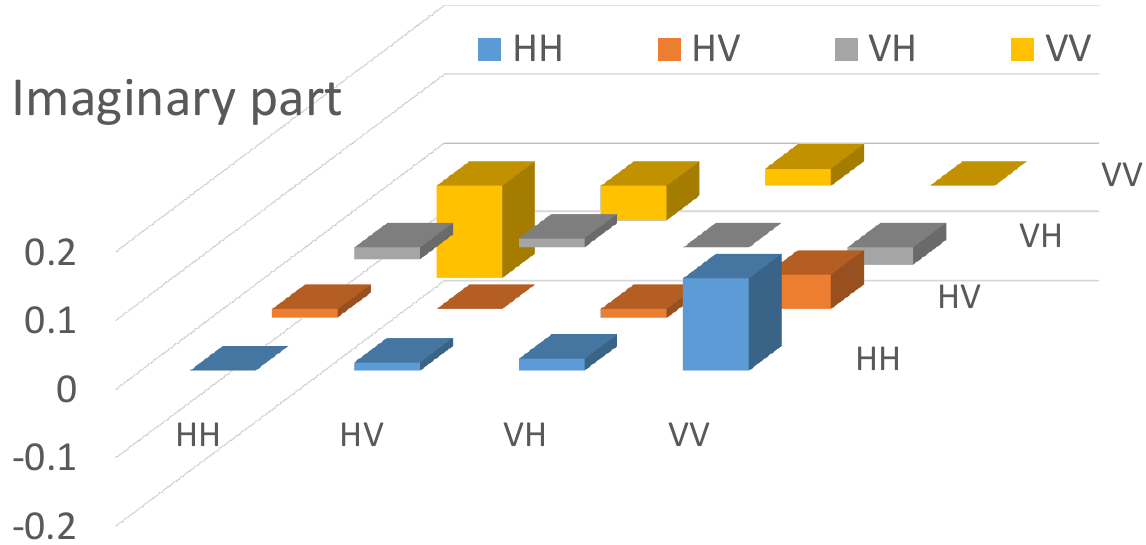}
	\caption{(Color online) Graphical representation of the real and imaginary parts of the maximum likelihood physical density matrix obtained from the experimental tomography data given in Tab.~(\ref{tabletomography}).}	
	\label{fig4}
\end{figure}

Fig.~(\ref{fig4}) shows the graphical representation of the real and imaginary parts of the maximum likelihood physical density matrix of our prepared entangled state, which is obtained by the experimental linear tomography from the experimental data given in Tab.~(\ref{tabletomography}). It also shows that our generated entangled state is not exactly the maximally entangled Bell state, but it is near to it, or is such a linear superposition of known Bell states. 

In the proceeding, we will calculate some important quantities derived from the density matrix such as Concurrence, the entanglement of formation, tangle, logarithmic negativity, and different entanglement entropies such as linear entropy, Von-Neumann entropy, and Renyi 2-entropy.

The Von Neumann entropy is an important measure of the purity of a quantum state which is given by \cite{kwiatstate}
\begin{eqnarray} \label{von1}
&& \mathcal{S}=-{\rm tr} (\hat \rho {\rm log_2} \hat \rho)=- \sum_{j=1}^{4} p_a^{(j)} {\rm log_2} p_a^{(j)},
\end{eqnarray}
where we have obtained $\mathcal{S}= 0.720 \pm 0.070 $. Note that $ \mathcal{S} $ for a pure state $ \vert AB \rangle $ ($ A,B=H,V $) is zero. Our obtained Von Neumann entropy shows the prepared state is the mixture of pure states $\rm HH,HV,VH,VV$.

The \textit{linear-entropy} for a two-qubit system which is given by \cite{kwiatstate}
\begin{eqnarray}
&& \mathcal{P}=\frac{4}{3} (1-{\rm tr} \hat \rho^2 )=\frac{4}{3} (1-\sum_{j=1}^{4}  p_a^{(j) 2} ),
\end{eqnarray}
is used to quantify the degree of mixture of a quantum state. Its value lies between zero and 1 ($ 0 \le \mathcal{P} \le 1$). We have obtained $\mathcal{P}= 0.425\pm0.040 $ for our entangled two photonic qubits, which shows our state is not pure state.

Let us now compute the \textit{quantum coherence} properties of a mixed quantum state for a two-qubit system, the concurrence, entanglement of formation, and tangle, which are equivalent measures of the entanglement of a mixed state. By considering the \textit{spin-flip} matrix $ \Sigma_f= \sigma_y \otimes \sigma_y $ in the pure computational basis, $ \vert AB \rangle $ ($ A,B=H,V $), as
\begin{eqnarray} \label{flip}
	\Sigma_f= \left( \begin{matrix}
	{0} & $ 0$ & $0$ & $-1$   \\
	$ 0$& $0$ & $1 $ & $0$ \\
	$0$ & $ 1 $ & $0$ & $0 $ \\
	$-1$ & $ 0 $ & $0$  & $ 0 $
	\end{matrix} \right) , \!\!\! \!\!\!\!\! \nonumber \\
\end{eqnarray}
one can define the \textit{non-Hermitian} matrix $ \hat R= \sqrt{\sqrt{\hat \rho} (\Sigma_f \hat \rho^\ast \Sigma_f) \sqrt{\hat \rho}} $ with the left and right eigenstates, $ \vert \xi_{L(R)}^{(a)} \rangle  $ and eigenvalues $ r_a $ with assumption $ r_1 \ge r_2 \ge r_3 \ge r_4 $. Thus, the concurrence is defined as \cite{kwiatstate}
\begin{eqnarray}
&& \mathcal{C}= {\rm Max} \left\lbrace 0 , \sum_{a=1}^{4} r_a {\rm sgn}(\frac{3}{2}-r_a) \right\rbrace , 
\end{eqnarray}
where $ \rm sgn (x>0)=1 $  and $ \rm sgn (x<0)=-1 $. For our case, the concurrence becomes $ \mathcal{C}= 0.602 \pm 0.051 $. 

The \textit{tangle} and the \textit{entanglement of formation } are, respectively, given by
\begin{eqnarray}
&& \mathcal{T}= \mathcal{C}^2, \\
&& \mathcal{E}= h (\frac{1+ \sqrt{1-\mathcal{C}^2}}{2}), 
\end{eqnarray}
where $ h(x)=-x {\rm log_2}x - (1-x) {\rm log_2}(1-x) $ is a monotonically increasing function. We found that $\mathcal{T}= 0.362 \pm 0.058 $ and $ \mathcal{E}=0.471 \pm 0.060 $.

Another measure to quantify the entanglement is \textit{Renyi 2-entropy}, which is defined for the density operator of a subsystem $ \hat \rho_A $ belonging to a bipartite pure state and can be defined as \cite{entropyreneyi}  
\begin{eqnarray}
&&  \Upsilon_A(\hat \rho)= - {\rm ln ~ tr} \hat \rho_A^2 , 
\end{eqnarray} 
where $ \hat \rho_A= {\rm tr_B} \hat \rho_{\rm AB} $ is the reducible density operator of the subsystem A. It can be easily shown that the reducible density operators can be expressed in terms of the element of bipartite density matrix as
\begin{eqnarray} \label{rhoAB}
\!\!\!\! \rho_A= \left( \begin{matrix}
{\rho_{11} +\rho_{33}} &  {\rho_{12} +\rho_{34}}  \\
{\rho_{12}^\ast +\rho_{34}^\ast} &  \rho_{22} +\rho_{44} 
\end{matrix} \right) , \quad   
\rho_B= \left( \begin{matrix}
{\rho_{11} +\rho_{22}} &  {\rho_{13} +\rho_{24}}  \\
{\rho_{13}^\ast +\rho_{24}^\ast} &  \rho_{33} +\rho_{44} 
\end{matrix} \right) .   \!\!\! \!\!\!\!\! \nonumber \\
\end{eqnarray}
The Renyi 2-entropy quantifies the purity of the subsystem A (B) of a pure state such that its high value shows a high degree of entanglement or low purity for the subsystem A(B). From our physical maximum likelihood density matrix, we find $ \Upsilon_A= 0.691 \pm 0.001 $, which shows our prepared two qubits state is not such a pure state.

Finally, we would like to calculate the \textit{Logarithmic negativity} as upper bound to the distillable entanglement which is derived from the Peres-Horodecki criterion for separability \cite{horodeckiSeparabale}. The negativity is defined as 
\begin{eqnarray} \label{negativity1}
&& E_{\mathcal{N}}(\rho)= {\rm log_2} {\rm tr} \sqrt{\hat \rho^{\dag \rm T_A} \hat \rho^{ \rm T_A}} ,
\end{eqnarray}
where $ \rm T_A $ is partial transpose respect to subsystem A. Note that the result of negativity is independent of the transposed party that because $ \rho^{ \rm T_A}= (\rho^{ \rm T_B})^{\rm T} $
It means that if the density matrix $ \hat \rho  $ is separable, then, all eigenvalues of $ \rho^{ \rm T_A} $ are non-negative. In other words, if the eigenvalues are negative; thus, $ \hat \rho $ is entangled. We find that $ E_{\mathcal{N}}(\rho)=0.678 \pm 0.042 $. Note that the full description of the error calculations can be seen in Ref.~\cite{kwiatstate}.


\section{summary and conclusion \label{summary}}

In the first part, we have presented a theoretical description with details to generate the maximally entangled state and its mathematical formulation using the type-I SPDC in pair-BBO crystal. We have also given the theory of the maximum likelihood density operator via the quantum state tomography, which these are useful for any graduate student in the laboratory. 

In the second part, we have described step-by-step to achieve experimentally an entangled quantum state close to the maximally polarization-entangled Bell state with high brightness. By measuring the CC rate, we have shown the strong violation from the classical physics or any HVT. Moreover, using the experimental tomographic CC-measurements together with the numerical optimization, i.e., the MLT, we have experimentally reconstructed the physical density matrix of the created entangled state. 
Furthermore, by obtaining the density matrix of the system, we have calculated different types of entanglement entropy such as linear entropy, Von-Neumann entropy and Renyi 2-entropy and also some other well-known entanglement measures in quantum information theory such as concurrence, the entanglement of formation, tangle, and logarithmic negativity.

It should be mentioned that the novelty of this work is the experimentally measurement of all entanglement entropies together the Bell inequalities which have been led to fully characterize our prepared entangled Bell state. Usually, all of these quantities are not be measured in an experiment. This helps young researchers in the Lab to better understand the Bell state properties. Moreover, unlike the most similar works in this context, we have presented and review a full theoretical description to better understand the experiment and measurements.

Finally, we hope that this experimental report which includes the full theoretical description to be useful for the graduate student or any researcher to start doing fundamental quantum optical experiment using the nonlinear quantum optics. Also, this high-brightness and low-rate photonic quantum source could be used for quantum measurements in short-range.

\section{AUTHOR CONTRIBUTIONS}
AMF organized the idea of this work. AMF and SAM performed the experiment, built the experimental setup, experimental measurements, adjustments, and analyzed the data. AMF and S.A.M. performed the theoretical calculation and numerical analysis, respectively. AMF and SAM both contributed to prepare the manuscript, but, AMF. wrote the manuscript. NSV discussed and commented about the experiment.

\section{acknowledgments}
All authors thank the ICQTs. Also, AMF and SAM thank Prof. Paul G. Kwiat and notably Christopher Karl Zeitler from university of illinois because of their useful comments in numerical part regarding the quantum state tomography.

\end{document}